\documentclass[]{spie}  

\usepackage{amsmath,amsfonts,amssymb}
\usepackage{graphicx}
\usepackage[numbers]{natbib}
\usepackage{comment}
\usepackage{wrapfig}
\usepackage{tabularx}
\usepackage{caption}
\usepackage[colorlinks=true, allcolors=blue]{hyperref}
\usepackage{aas_macros}
\usepackage{xspace}
\usepackage{lineno}

\newcommand{\registered}{\textsuperscript{\textregistered}\xspace}

\usepackage{tikz,xcolor,hyperref}
\definecolor{lime}{HTML}{A6CE39}
\DeclareRobustCommand{\orcidicon}{%
    \begin{tikzpicture}
    \draw[lime, fill=lime] (0,0) 
    circle [radius=0.16] 
    node[white] {{\fontfamily{qag}\selectfont \tiny ID}};
    \draw[white, fill=white] (-0.0625,0.095) 
    circle [radius=0.007];
    \end{tikzpicture}
    \hspace{-2mm}
}
\newcommand{\orcid}[1]{\href{https://orcid.org/#1}{\orcidicon}}

\title{Investigation and Mitigation of a Prominent Off-Axis Stray Light Path in Rubin Observatory Commissioning}

\author[1,2,3,4]{Alex~Drlica-Wagner\orcid{0000-0001-8251-933X}}
\author[5,6]{Alessio~Taranto\orcid{0009-0009-3271-3498}}
\author[5]{Gabriele~Rodeghiero\orcid{0000-0002-3469-9863}}
\author[7]{Joshua~E.~Meyers\orcid{0000-0002-2308-4230}}
\author[8]{John~Andrew}
\author[8]{Douglas~R.~Neill}
\author[8]{Christian~Aguilar}
\author[8]{Brian~Stalder\orcid{0000-0003-0973-4900}}
\author[9]{Robert~H.~Lupton\orcid{0000-0003-1666-0962}}
\author[3,10]{Aashay~Pai\orcid{0009-0008-9641-6065}}
\author[9]{Lee~S.~Kelvin\orcid{0000-0001-9395-4759}}
\author[11]{Aaron~E.~Watkins\orcid{0000-0003-4859-3290}}
\author[6,5]{Luca~Rosignoli\orcid{0000-0002-0327-5929}}
\author[12]{Hannah~M.~M.~Pollek}
\author[8]{Anastasia~Alexov\orcid{0009-0000-7835-3963}}

\affil[1]{\small Fermi National Accelerator Laboratory, Batavia, IL 60510, USA}
\affil[2]{\small Department of Astronomy \& Astrophysics, University of Chicago, Chicago, IL 60637, USA}
\affil[3]{\small Kavli Institute of Cosmological Physics, University of Chicago, Chicago, IL 60637, USA}
\affil[4]{\small NSF-Simons AI Institute for the Sky (SkAI),172 E. Chestnut St., Chicago, IL 60611, USA}
\affil[5]{\small INAF Osservatorio di Astrofisica e Scienza dello Spazio Bologna, Via P. Gobetti 93/3, 40129, Bologna, Italy}
\affil[6]{\small Department of Physics and Astronomy (DIFA), University of Bologna, Via P. Gobetti 93/2, 40129, Bologna, Italy}
\affil[7]{\small Kavli Institute for Particle Astrophysics and Cosmology, SLAC National Accelerator Laboratory, 2575 Sand Hill Rd., Menlo Park, CA 94025, USA}
\affil[8]{\small NSF-DOE Vera C.\ Rubin Observatory / NSF NOIRLab, 950 N.\ Cherry Ave., Tucson, AZ  85719, USA}
\affil[9]{\small Department of Astrophysical Sciences, Princeton University, Princeton, NJ 08544, USA}
\affil[10]{\small Department of Physics, University of Chicago, Chicago, IL 60637, USA}
\affil[11]{\small Centre for Astrophysics Research, University of Hertfordshire, Hatfield, Hertfordshire, AL10 9AB, UK}
\affil[12]{SLAC National Accelerator Laboratory, 2575 Sand Hill Rd., Menlo Park, CA 94025, USA}

\authorinfo{Further author information: Alex Drlica-Wagner (kadrlica@fnal.gov)}

\begin{document} 
\maketitle

\begin{abstract}
The ``scratched tape'' stray light feature is the most prominent and prevalent stray light artifact identified during the commissioning of the Vera C.\ Rubin Observatory. The scratched tape feature originates when light from large off-axis angles ($\sim$20\,deg) passes between the mid-level and center-section light baffles, reflects off the primary mirror (M1), and illuminates the LSST Camera focal plane. This scenario represented an unobstructed stray light path to the sky during Rubin commissioning due to delays in the integration of the dome slit light--wind screen.  This document describes the identification, modeling, characterization, and mitigation of the scratched tape stray light artifact.
\end{abstract}


\section{Introduction}
\label{sec:introduction}

\begin{figure*}[t!]
    \includegraphics[width=0.50\textwidth, clip, trim={0cm 0.75cm 0cm 2.0cm}]{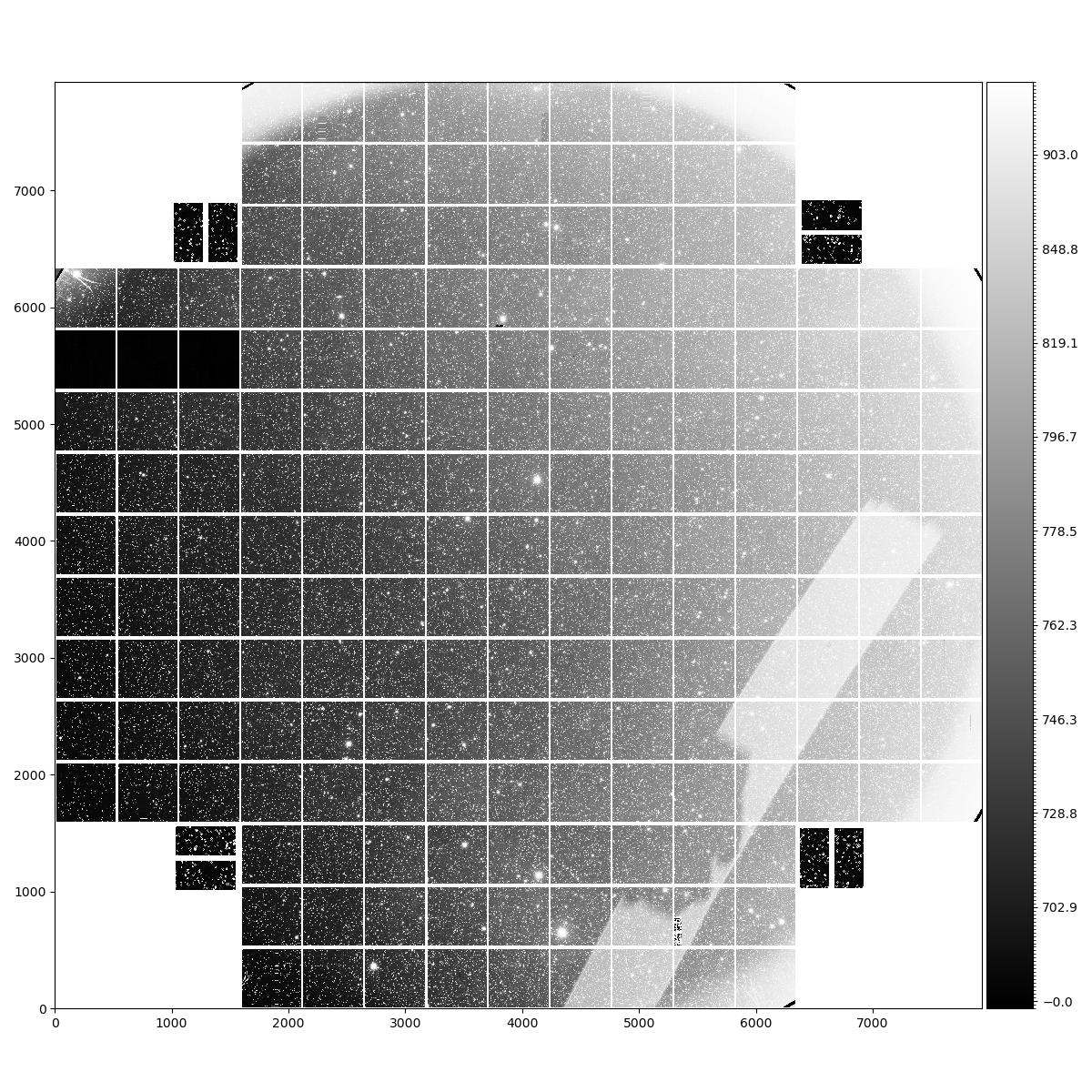}
    \includegraphics[width=0.49\textwidth, clip, trim={0.5cm 0cm 0cm 1.5cm}]{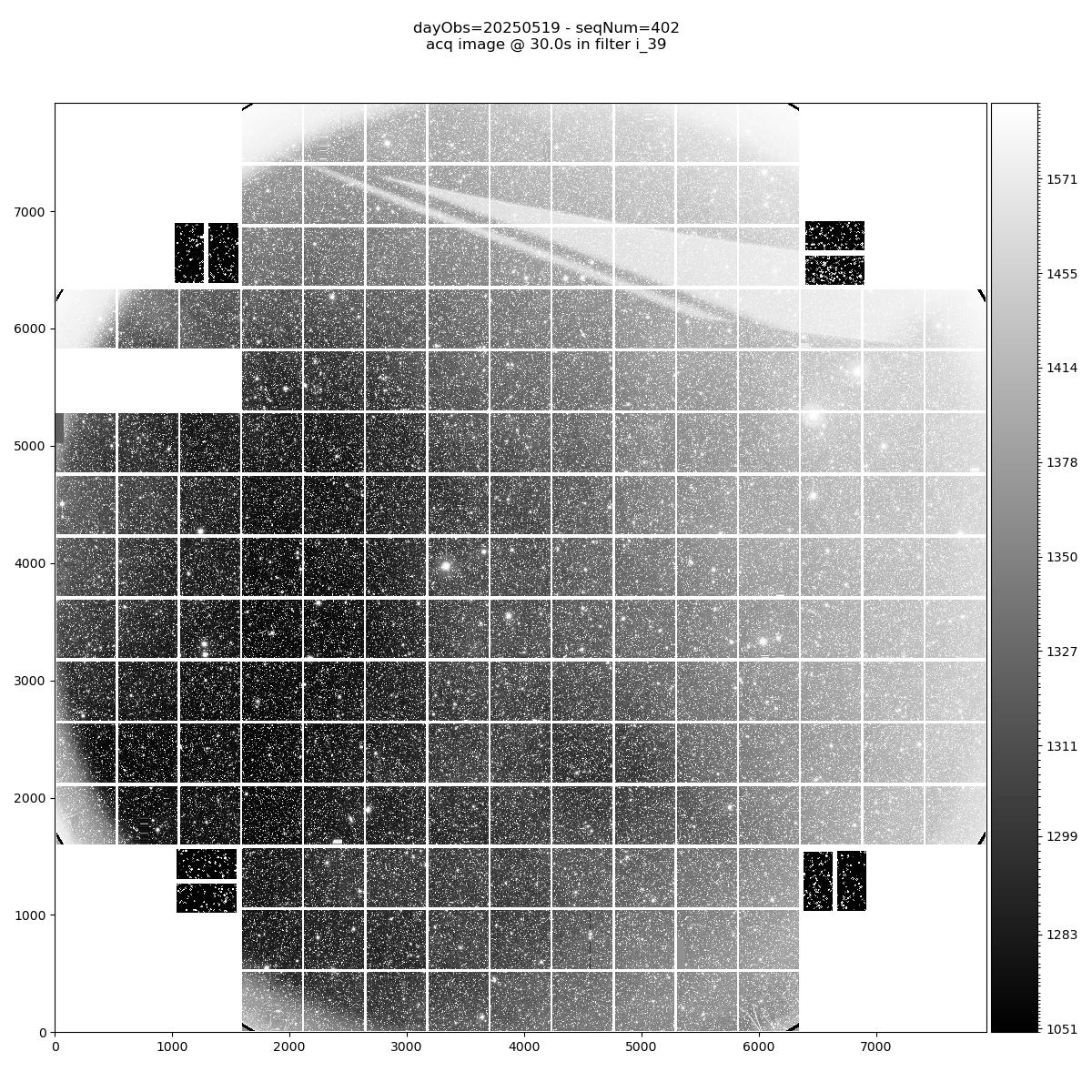}
    \caption{\label{fig:tape} Examples of stray light features associated with the ``scratched tape'' stray light path in LSSTCam images from Rubin commissioning. These examples originate from $\alpha$\,Cen ({\it V}~$\sim$~$-$0.29\,mag), which is located $\sim$21\,deg off-axis and illuminates the primary mirror (M1) through a gap between the mid-level and center-section light baffles. (Left) Prototypical ``scratched tape'' morphology (visit = 2025050500421; 15\,s exposure in $r$-band). (Right) Prototypical ``sail'' morphology coming from a similar light path (visit = 2025051900402; 30\,s exposure in $i$-band). The difference in morphology comes from physical obscurations on the TMA (see Figure~\ref{fig:pinhole}).}
\end{figure*}

The unique, wide-field design of the Simonyi Survey Telescope at the Vera C.\ Rubin Observatory makes it particularly susceptible to stray and scattered light. Since one of the primary goals of Rubin is to investigate the low-surface-brightness Universe \citep{2019ApJ...873..111I}, it is particularly important to identify, model, and mitigate stray-light artifacts. In particular, the delayed installation of the Rubin dome slit light--wind screen (LWS; \citep{Marchiori:2024}) has led to a number of unanticipated stray light features that were identified during Rubin commissioning \citep{Rodeghiero:2026}. Here, we focus specifically on efforts related to the ``scratched tape'' stray light feature (Figure~\ref{fig:tape}), which is the most prominent and prevalent stray light artifact identified during Rubin commissioning \citep{SITCOMTN-166}. We describe the discovery, origin, characterization, and successful mitigation of this stray light artifact.

The scratched tape artifact is a wide-area stray light feature that is visually apparent in $>$\,5\% of images collected by the LSST Camera (LSSTCam) during the first year of on-sky commissioning and early operations \cite{Rodeghiero:2026}. As described in Section~\ref{sec:cause}, the scratched tape was found to originate from an unobstructed light path to the sky that passes between the mid-level and center-section light baffles on the Telescope Mount Assembly (TMA) of the Simonyi Survey Telescope \citep{2022SPIE12182E..0WT}. The forthcoming LWS is expected to block this path to the sky once it is installed and operating.  However, the partially completed state of the Rubin dome during commissioning, science validation, and early operations made it possible for astronomical sources located $\sim$20\,deg off-axis to illuminate the primary mirror (M1) and have their light reflected directly into LSSTCam---i.e., bypassing the secondary and tertiary mirrors (M2 and M3, respectively). 
The result is a sharp, several-degree-long, rectilinear feature with surface brightness that has been measured to be $\sim$\,20\% of the dark-sky background in bluer bands when sourced by bright stars (Figure~\ref{fig:tape}).
The scratched tape feature manifests in a variety of morphologies, brightnesses, and prominences, and several morphologically similar artifacts (e.g., the ``pillow'', ``muddy shoe'', etc.; \citep{Rodeghiero:2026}) are suspected to be associated with the same light path. In this document, we use the term ``scratched tape'' to collectively refer to all features that arise from the same off-axis stray light path.
Figure~\ref{fig:tape} shows two common morphologies of the scratched tape feature sourced by the bright star system $\alpha$\,Centauri ($\alpha$\,Cen; $V \sim -0.29$\,mag \cite{1982bsc..book.....H})\footnote{The combined $V$-band magnitude is computed from $\alpha$\,Cen A ($V = -0.01$) and $\alpha$\,Cen B ($V = +1.33$) \cite{1982bsc..book.....H}.} when it was located $\sim$21\,deg off-axis from the telescope boresight.

\section{Identification}
\label{sec:identification}

The scratched tape artifact was identified in early commissioning images from LSSTCam (the first documented instance was on 17 April 2025). The origin of this feature was initially unclear, although it was found to consistently impact observations of specific commissioning fields (i.e., SV\_225\_-40). A visual inspection campaign was mounted to catalog and characterize the appearance of this and other stray-light features \citep{Rodeghiero:2026}. By correlating the appearance of the scratched tape with the locations of bright stars, it was determined that some of the brightest instances of the scratched tape appeared in observations taken when $\alpha$\,Cen was located $\sim$\,21\,deg off-axis. Based on this analysis, observations were planned and executed to reproduce the scratched-tape feature on 28 May 2025. While the ability to successfully reproduce the feature strongly indicated that $\alpha$\,Cen was the light source, the exact off-axis light path responsible for this feature was still unclear. In the meantime, it was found that $>$\,5\% of images were visually affected by similar artifacts that spanned a wide range of morphologies and brightnesses, and it was suspected that a much larger fraction of images were affected at a lower level that was not immediately flagged by visual inspection.

\begin{figure*}[t!]
    \centering
    \includegraphics[width=0.75\textwidth]{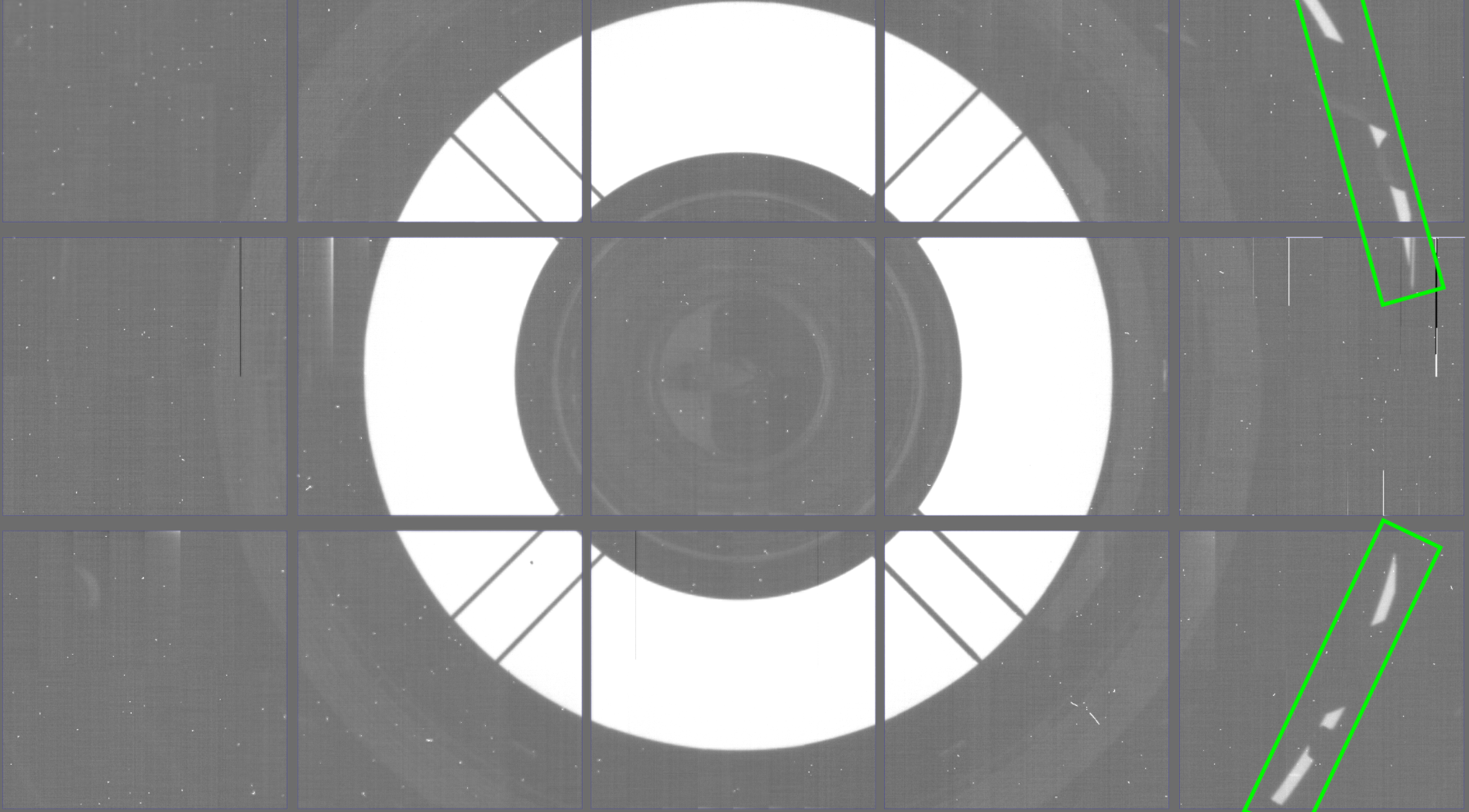}\\[+0.5em]
    \includegraphics[width=0.75\textwidth]{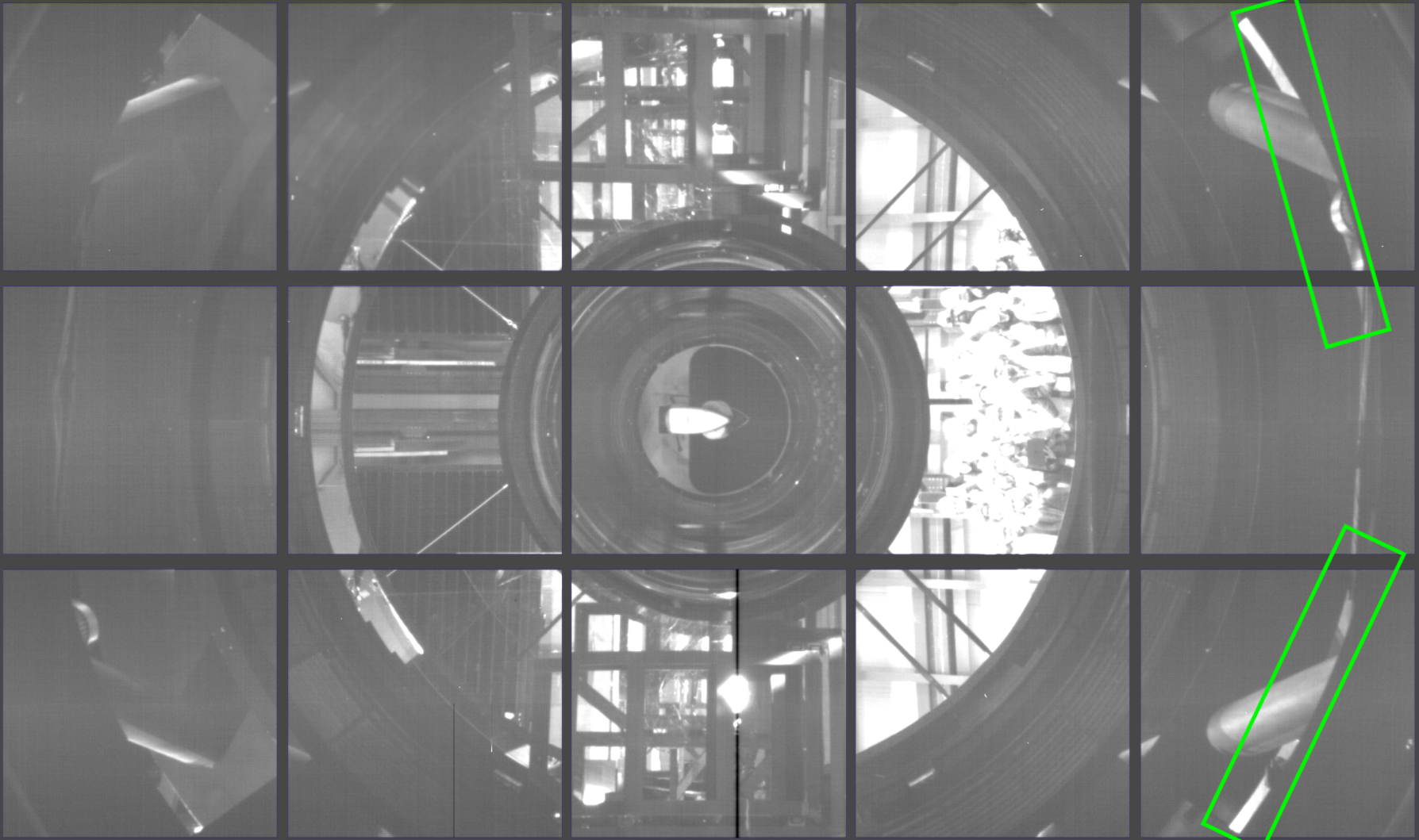} 
    \vspace{1em}
    \caption{\label{fig:pinhole} Images taken with the pinhole filter showing the gap between the mid-level and center-section light baffles that is responsible for the scratched tape feature (green rectangles). (Top) Twilight flat (visit = 2025070100103; 120\,s exposure) showing scratched-tape-like morphology at large incident angles. (Bottom) In-dome image at higher illumination level (visit = 2025070200040; 1\,s exposure) showing the scratched tape path in the context of the TMA geometry. 
    }
\end{figure*}

\section{Physical Origin}
\label{sec:cause}

\begin{figure*}[t!]
    \centering
    \includegraphics[width=0.43\textwidth]{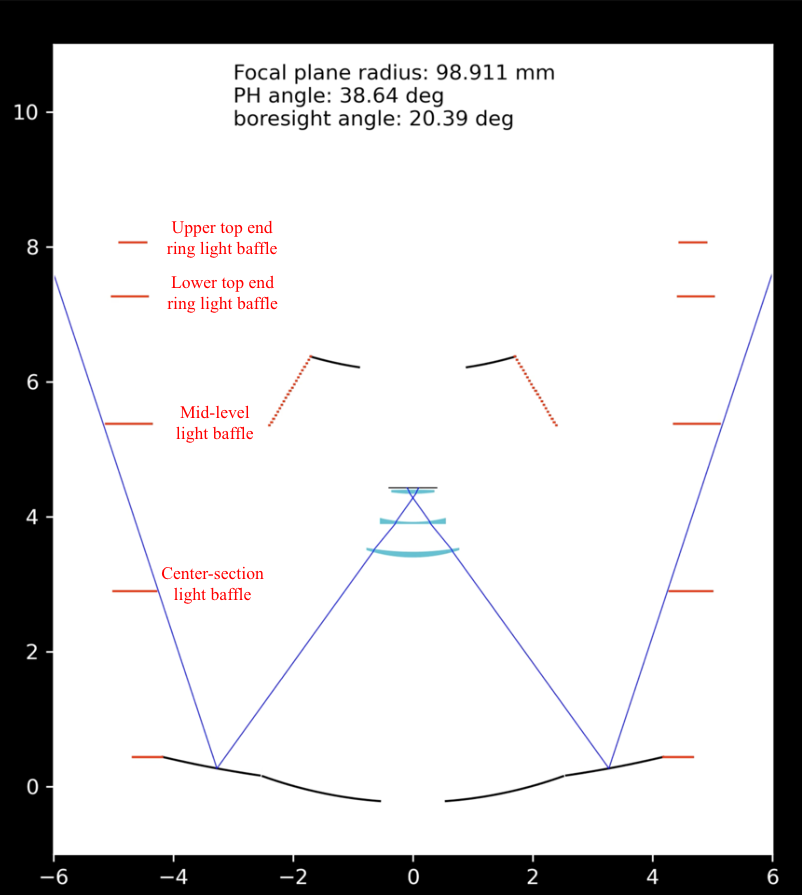}
    \includegraphics[width=0.56\textwidth, trim = {3.5cm 0cm 3cm 0cm}, clip]{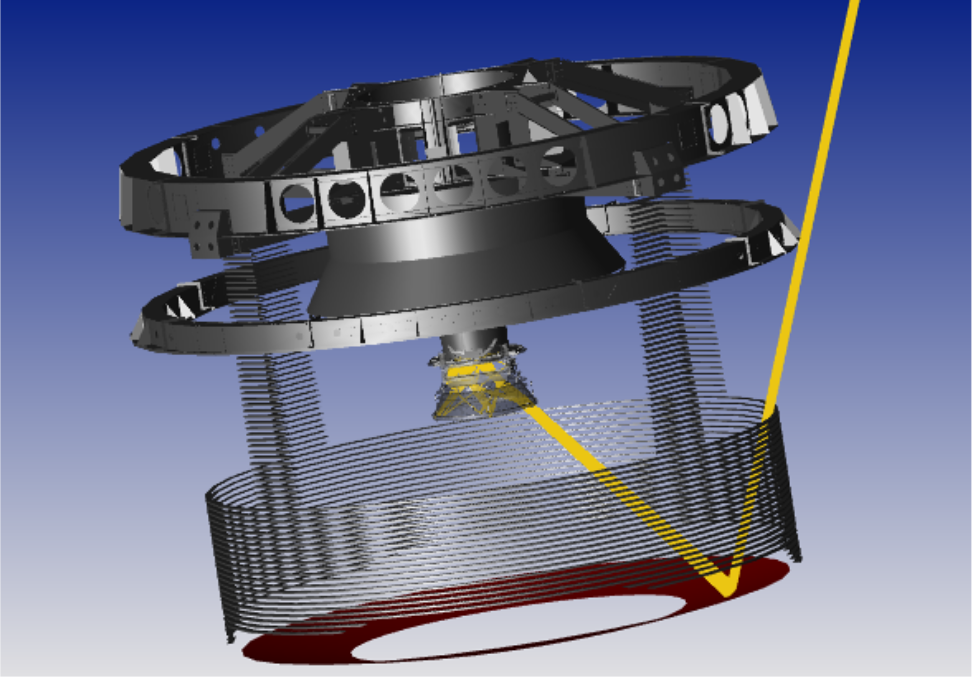}
    \vspace{1em}
    \caption{\label{fig:raytrace} Ray tracing models showing the scratched tape light path between the mid-level and center-section light baffles. (Left) The \texttt{batoid} model showing a ray passing through the scratched tape gap and into the center pinhole. The projected radius on the focal plane (98.911\,mm) corresponds to the position of the scratched tape seen in pinhole twilight flats. Baffles (red lines) are labeled following the nomenclature of TMA engineering drawings \citep{LTS-213}. (Right) A \texttt{Zemax} model showing the scratched tape light path with a more complete model of the TMA and LSSTCam. \texttt{Zemax} modeling indicated that it is possible for photons to reach the detectors for off-axis angles of 19.8 $< \theta <$ 26\,deg due to secondary bounces off surfaces within the camera.}
\end{figure*}

After demonstrating the ability to reproduce the scratched tape feature on-sky, emphasis was placed on understanding the light path that allowed a bright star located $>$\,20\,deg off-axis to impact LSSTCam images. Investigation of images taken with the pinhole filter during twilight in early May 2025 proved extremely useful (Figure~\ref{fig:pinhole}). The pinhole filter is an opaque mask with a configurable number of small pinholes that can be inserted directly into the LSSTCam filter changer to produce an image of the telescope pupil, including the mechanical structures that surround the system clear aperture. We identified a direct connection between the appearance of the scratched-tape features observed on-sky and the light patterns revealed by the pinhole images. Morphologically similar bright features were observed at large angles from the central pinhole in twilight images. Additional pinhole observations were executed in July 2025, including both on-sky twilight images and illuminated in-dome images. Investigation of these images identified the stray light path as a gap between the mid-level and center-section light baffles (Figures~\ref{fig:pinhole} and \ref{fig:raytrace}). The existence of this gap was experimentally verified by directed illumination with the collimated beam projector \citep{2016SPIE.9910E..0VC, Mondrik:2023, SITCOMTN-152} at large off-axis angles on 3 September 2025.
In addition, models of the telescope and optical system with the \texttt{batoid} ray-tracing software \citep{batoid} and Ansys Zemax OpticStudio\registered (\texttt{Zemax}) \citep{zemax} were able to reproduce this stray light path to the sky (Figure~\ref{fig:raytrace}). 

\begin{figure*}[t!]
    \centering
    \includegraphics[width=0.55\textwidth]{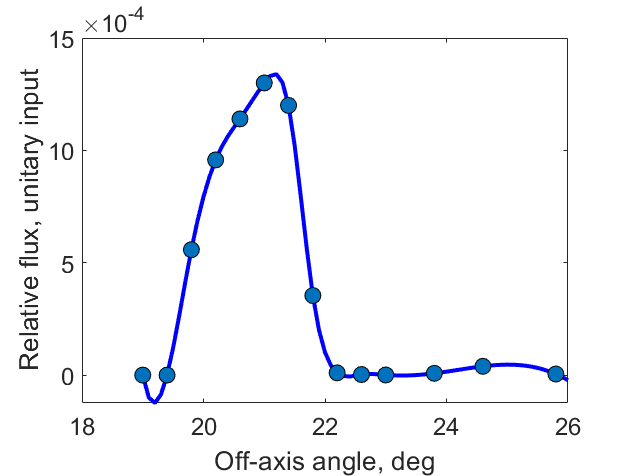}
    \includegraphics[width=0.42\textwidth]{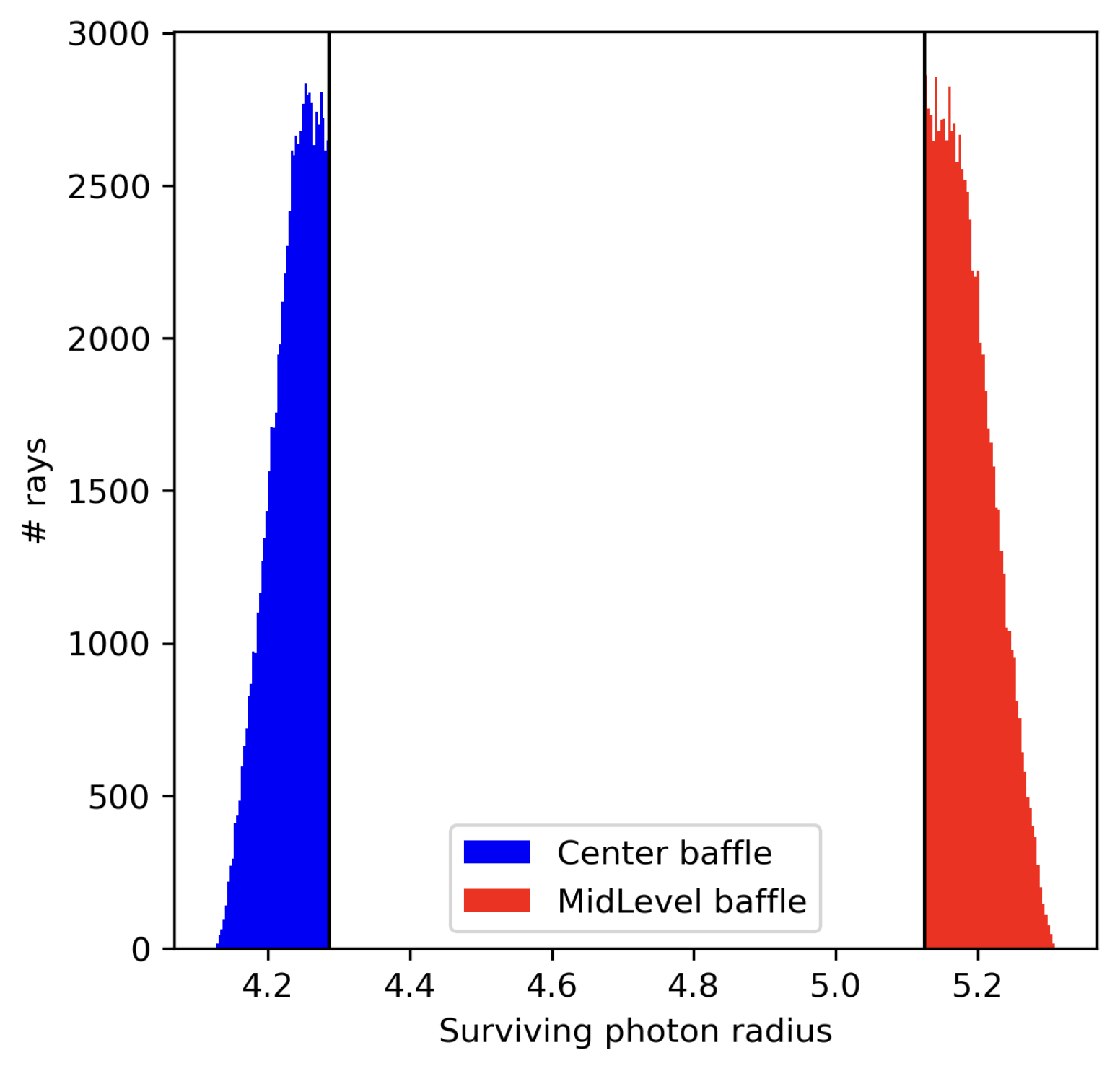}
    \vspace{1em}
    \caption{\label{fig:st_flux}
    (Left) Ray-tracing simulations with \texttt{Zemax} show the relative flux on the LSSTCam focal plane via the scratched tape light path for a unitary flux injected into the telescope pupil. 
    The relative flux reaching the LSSTCam focal plane peaks at $\sim$\,10$^{-3}$ and falls to $\sim$\,10$^{-6}$ at off-axis angles $\gtrsim$\,22\,deg.
    (Right) Ray-tracing simulations run with \texttt{batoid} show the distribution of $\sim 2\times 10^7$ photons fired in the direction of the scratched tape gap between the mid-level and center-section light baffles. The simulation tracks the rays that survive the trip through the baffles, reflect off M1, enter LSSTCam, and reach the detectors. The radial distribution of the surviving rays at the heights of the mid-level baffle and the center-section baffle are shown in red and blue, respectively. This study showed that the mid-level baffle would need to be extended outward by 185\,mm to block all surviving rays. However, these simulations used the TMA design dimensions \cite{LTS-213} and do not account for the 10\,mm uncertainty on the as-built TMA.
}
\end{figure*}

Stray light passing between the mid-level and center-section baffles can contaminate the LSSTCam focal plane across a wide range of off-axis separations and clocking angles. Based on non-sequential \texttt{Zemax} modeling that included a representative mechanical model of the TMA baffles and structure, the full range of the incident off-axis separation angle from the boresight, $\theta$, was determined to be 19.8 $ \lesssim \theta \lesssim$ 26\,deg (Figure~\ref{fig:st_flux}). The \texttt{Zemax} modeling suggested that the angular range for bright scratched-tape features is concentrated at 20 $\lesssim \theta \lesssim$ 22\,deg, where the illumination can reach $\gtrsim$10$^{-3}$ of the flux injected into the telescope pupil. Beyond 22\,deg, the illumination comes from secondary bounces off structures internal to the camera, and the relative flux deposited onto the LSSTCam focal plane falls to $\sim$10$^{-6}$. In terms of off-axis clocking angle acceptance, the gap between the mid-level and center-section baffles exists in each quadrant of the TMA. However, the dome slit generally blocks off-axis light from the sky in the azimuth direction (i.e., the $\pm$X quadrants in TMA coordinates). Thus, the scratched tape feature is observed predominantly when bright objects are offset in the altitude direction (i.e., the $\pm$Y quadrants in TMA coordinates). Modeling shows that the LWS is expected to block the scratched tape stray light path in the altitude direction once it is fully installed and operating.

\section{Characterization}
\label{sec:brightness}

Extensive visual inspection has produced a large database of images affected by scratched tape and other stray light artifacts \citep{Rodeghiero:2026}. During the period from 17 April 2025 to 3 July 2025, there were 1,093 documented examples of the scratched tape artifact out of 22,308 exposures ($\sim$4.9\%). However, much of the observing during commissioning was highly biased toward a small number of fields, and it can be argued that a more useful metric is the number of \emph{fields} in which the scratched tape occurs. Grouping visits with boresight pointing (R.A., Dec.) within 0.2\,deg of each other, we found 2,998 unique sky configurations, of which 188 show the scratched-tape artifact ($\sim$6.3\%). 
However, early visual inspection campaigns classified the scratched tape into several distinct features that were only later determined to result from the same stray light path. 
In reality, the scratched tape can manifest in a wide range of morphologies, sizes, and brightnesses due to a variety of factors (e.g., the brightness and location of the light source, environmental conditions, etc.).
Since the above analysis was performed only for features that were classified with the prototypical scratched tape morphology (as shown in the left panel of Figure~\ref{fig:tape}), it underestimates the true incidence of related features.  Furthermore, as flat fielding improved during commissioning, it became possible to identify fainter instances of the scratched tape artifact. 
Thus, this analysis can be taken as a lower limit on the true occurrence rate of the scratched tape, which we conservatively quote as visibly affecting $\gtrsim\,5\%$ of LSSTCam images during the first year of on-sky data taking. This number agrees with later estimates shown in Section~\ref{sec:results} and Rodeghiero et al.\ (2026) \cite{Rodeghiero:2026}.

A devoted analysis of the flux and surface brightness of the scratched tape features originating specifically from $\alpha$\,Cen was performed using 10 distinct instances per band. This analysis involved identifying the bounding region of the scratched tape and obtaining the mean calibrated flux per pixel and surface brightness.
In order to estimate the surface brightness of the feature, it is necessary to model and subtract a smooth sky background component. Estimating the background on small spatial scales, as is done in the standard LSST data processing pipeline, results in a loss of intensity because light from the scratched-tape feature is often included in the background model. To avoid this misinterpretation, we performed our analysis on images that have had instrument signatures removed (e.g., bias, dark, and flat-field corrections applied) but have not been background subtracted (i.e., the \texttt{post\_isr\_image} data products). We then estimate the background over the entire focal plane using a low-order polynomial fit to avoid removing light associated with the scratched tape feature.
The scratched tape feature associated with $\alpha$\,Cen was characterized in 10 images in each band, and the mean surface brightness and standard deviations are reported in Table~\ref{tab:sb}.

\begin{table}[t!]
\vspace{1em}
\centering
\begin{tabular}{c c c}
\hline
Band & Scratched Tape     & Dark-Sky Background \\
     & (mag/arcsec$^2$)   & (mag/arcsec$^2$) \\
\hline
\hline
 $u$ & $24.65\,\pm\,0.13$ & 23.05 \\
 $g$ & $23.97\,\pm\,0.16$ & 22.25 \\
 $r$ & $23.01\,\pm\,0.27$ & 21.20 \\
 $i$ & $22.93\,\pm\,0.15$ & 20.46 \\
 $z$ & $22.79\,\pm\,0.16$ & 19.61 \\
 $y$ & $21.53\,\pm\,0.66$ & 18.60 \\
\hline
\end{tabular}
\vspace{1em}
\caption{Surface brightness of the scratched tape features originating from $\alpha$\,Cen computed from 10 images. The reference dark-sky background for LSST is taken from SMTN-002 \citep{SMTN-002}. \label{tab:sb}}
\end{table}

Note that these surface brightness estimates were assembled from observations collected over several nights when the scratched tape feature could be associated with $\alpha$\,Cen. The reported standard deviation includes variations in the observing conditions, differences in the precise location of $\alpha$\,Cen with respect to the gap between the mid-level and center-section light baffles, and the exact geometry of the scratched tape feature on the focal plane. Furthermore, while the wavelength dependence of the scratched tape in this analysis is expected to follow the spectral energy distribution of $\alpha$\,Cen, this has not been explicitly verified.
Despite these caveats, this analysis suggests that the scratched tape feature originating from bright stars can contribute a flux per pixel that is comparable to $\sim$\,20\% of the nominal dark-sky background in the bluer bands \citep{SMTN-002}. 
Furthermore, we find that the total integrated flux of the scratched tape feature is ${\sim}\,1.3 \times 10^{-3}$ of the flux expected from direct on-axis illumination by $\alpha$\,Cen, which roughly agrees with the relative flux predicted by the \texttt{Zemax} model (Figure~\ref{fig:st_flux}).

While the analysis in this section focuses specifically on scratched tape features originating from $\alpha$\,Cen, the scratched tape feature will occur for {\it any} light source that is offset from the boresight by an appropriate altitude and azimuth. Only bright sources (e.g., bright stars, planets, the Moon, the twilight sky, etc.) produce scratched tape features that are easily identified in the LSSTCam images; however, there are indications that fields with many repeated observations taken with similar telescope altitude and azimuth (i.e., commissioning data for the Deep Drilling Fields) may have lower surface-brightness scratched tape features that become apparent after image coaddition. Together, these characterization studies presented a strong motivation for blocking the scratched tape stray light path before the start of LSST.

\section{Mid-Level Baffle Extension}
\label{sec:mitigation}

\begin{figure*}[t!]
\centering
\begin{minipage}[t]{0.45\textwidth}
\vspace{0pt}
\centering
\includegraphics[width=\textwidth, trim={2cm 0cm 0cm 0cm}, clip]{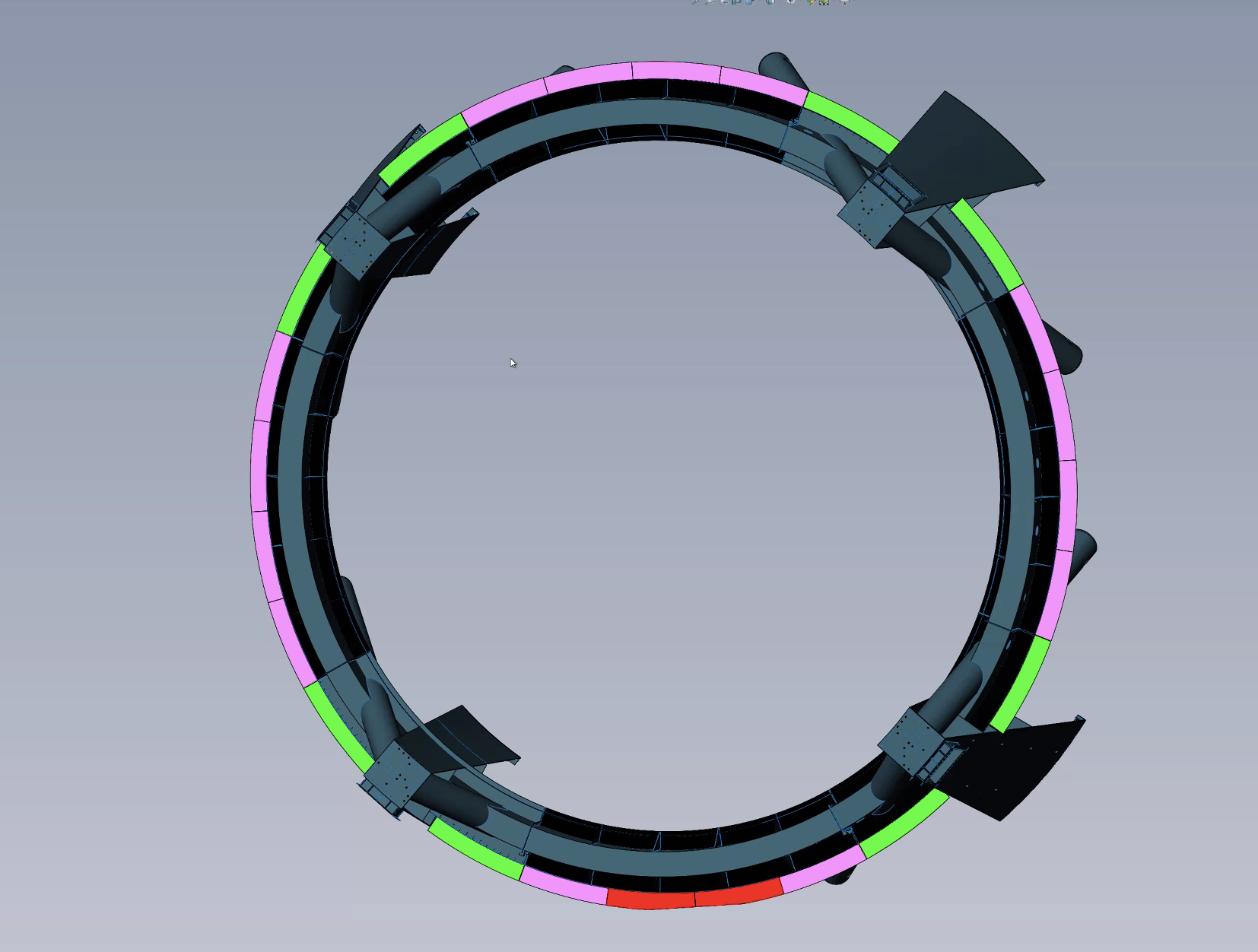}
\vspace{0em}
\captionof{figure}{Drawing of the 220\,mm outward extension to the mid-level light baffle to block the scratched tape light path. The mid-level baffle extension is composed of 24 panels that were installed incrementally (colored green, pink, and red).\label{fig:baffle}}
\end{minipage}
\hfill
\begin{minipage}[t]{0.5\textwidth}
\vspace{1cm}
\centering
\begin{tabularx}{\textwidth}{@{} l >{\centering\arraybackslash}X c @{}}
\hline
Date & Baffle Extension Component & Completion Status  \\
\hline\hline
2026-02-19 & $-$Y all panels   & {\bf (6/6)} \\
2026-02-20 & $+$X lower panels & (4/6) \\
2026-02-24 & $-$X lower panel  & (1/6) \\
2026-02-25 & $-$X lower panel  & (2/6) \\
2026-02-26 & $+$Y outer panels & (2/6) \\
2026-02-27 & $-$X upper panels & {\bf (6/6)} \\
2026-03-04 & $+$Y inner panels & {\bf (6/6)} \\
2026-03-06 & $+$X upper panels & {\bf (6/6)} \\
2026-04-02 & Corner overlap panels & {\bf (8/8)} \\
\hline
\end{tabularx}
\vspace{1.5em}
\captionof{table}{Mid-level baffle extension installation timeline. The completion status gives the cumulative number of panels installed in that quadrant after the listed date. \label{tab:timeline}}
\end{minipage}

\vspace{1em}

\includegraphics[width=0.65\textwidth]{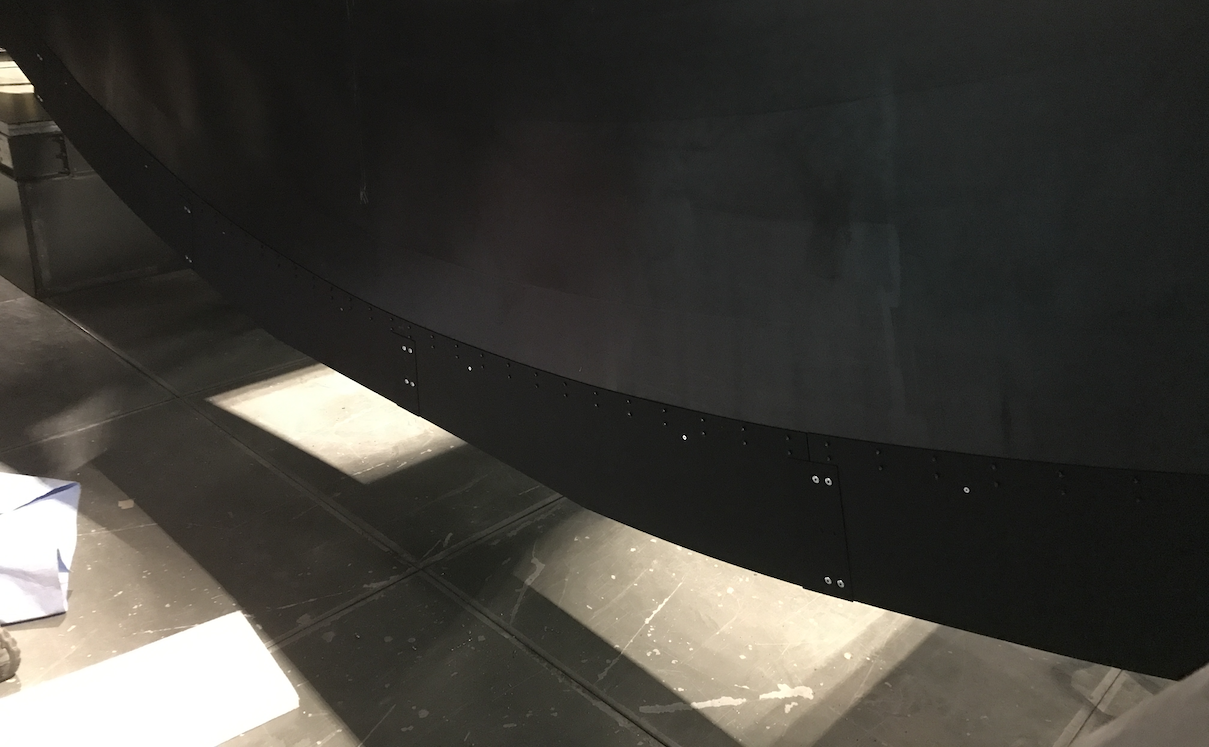}
\includegraphics[width=0.30\textwidth]{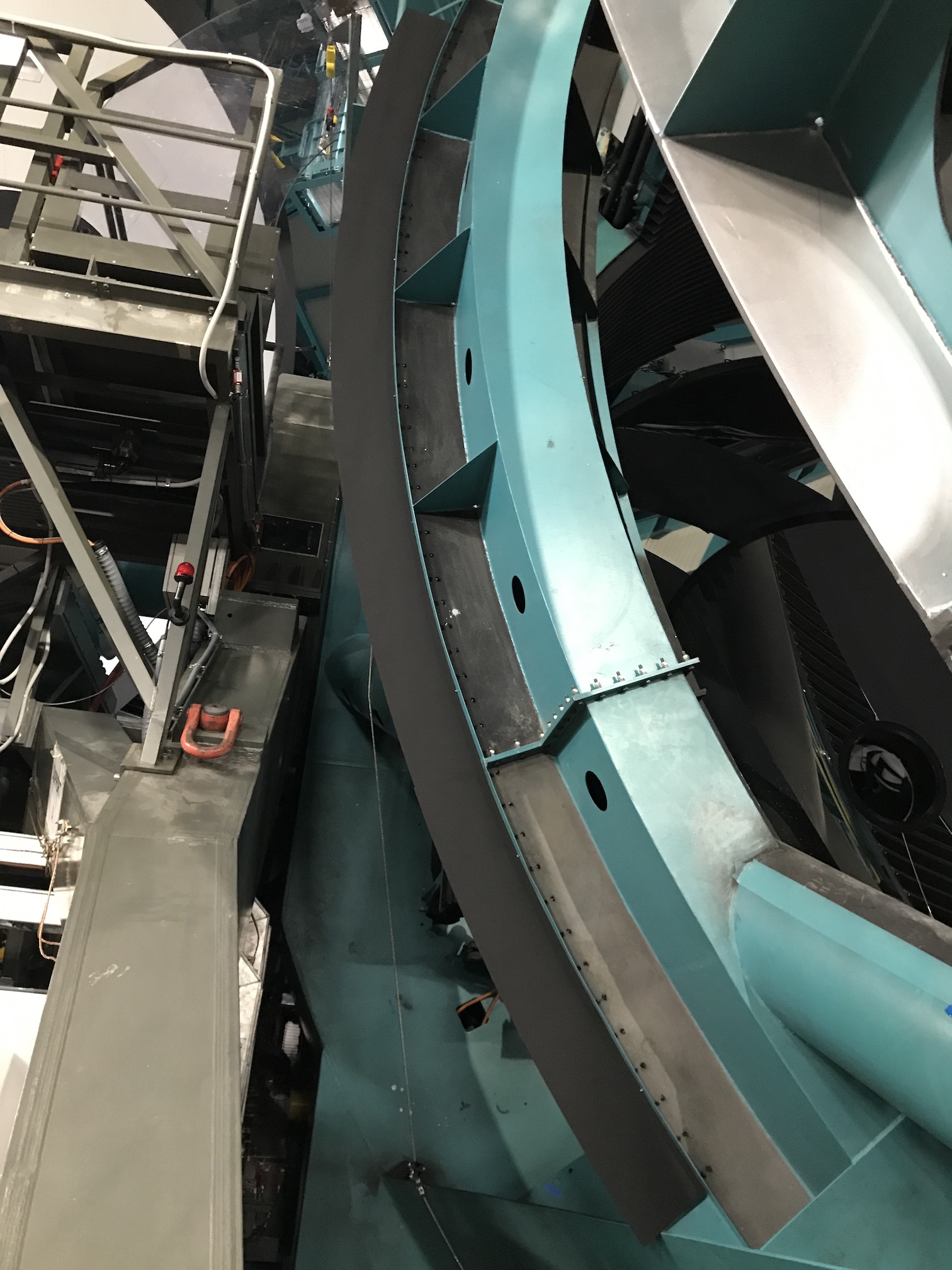}
\vspace{1em}
\caption{\label{fig:install} Mid-level light baffle extension panels installed on the $-$Y quadrant (left) and $+$X quadrant (right) of the TMA. The baffle extension panels are the darker material attached to the outer edge of the mid-level ring.
}
\end{figure*}

The ray tracing studies described in Section~\ref{sec:cause} indicated that the scratched tape light path could be effectively mitigated by extending the mid-level light baffle outward (Figure~\ref{fig:baffle}). This mitigation strategy has no impact on the clear aperture of the telescope.
Ray-tracing simulation with \texttt{batoid} modeled the paths of ${\sim}$\,2 $\times$ 10$^7$ photons fired in the direction of the scratched tape gap between the mid-level and center-section light baffles. This simulation tracked rays that survived the trip through the baffles, reflected off M1, entered LSSTCam, and reached the detectors. It was found that extending the mid-level baffle outward by 185\,mm would block all simulated photons from reaching the detectors. However, these simulations used the design dimensions of the TMA \cite{LTS-213}, did not account for uncertainty on the positions/dimensions of the as-built TMA, and did not account for reflections off components inside the camera.
Further modeling with \texttt{Zemax} showed that an extension of 200\,mm would effectively block the primary scratched tape component out to $\sim$\,23\,deg. A small fraction of photons originating with off-axis angles $>$\,23\,deg can enter the camera and reflect off components internal to the camera (e.g., the L2 holder pads, auto-changer, etc.) to reach the focal plane. However, the flux of these reflections is suppressed by a factor of $\sim$10$^{-3}$ relative to the scratched tape itself (Figure~\ref{fig:st_flux}).
Given the tolerances on all large TMA components ($\sim$\,10\,mm) and allowing for some margin of error, the recommended width of the mid-level baffle extension was 220\,mm.

\begin{figure*}[t!]
    \center
    \includegraphics[width=0.49\textwidth, trim={0cm 0cm 0cm 2.5cm}, clip]{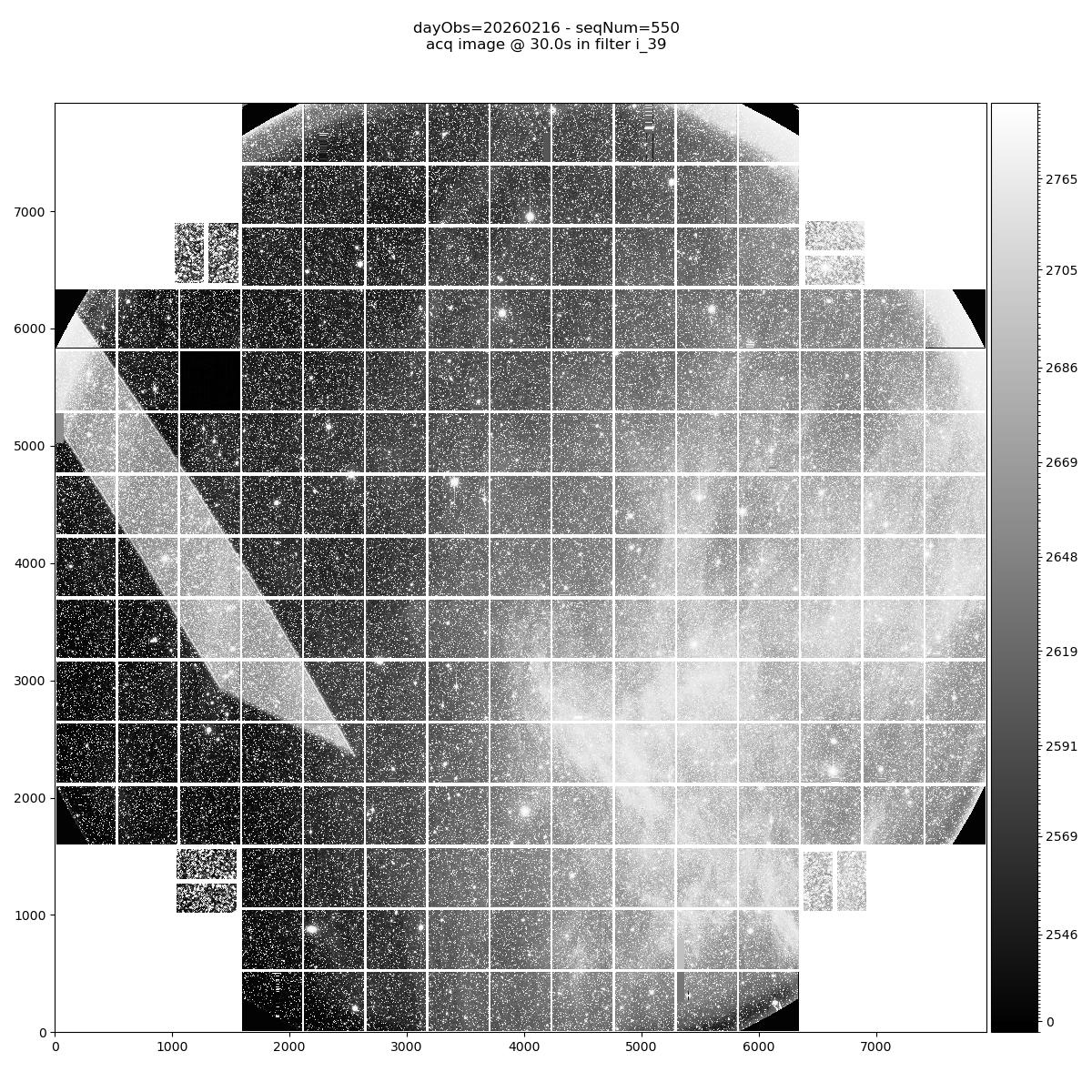}
    \includegraphics[width=0.49\textwidth, trim={0cm 0cm 0cm 2.5cm}, clip]{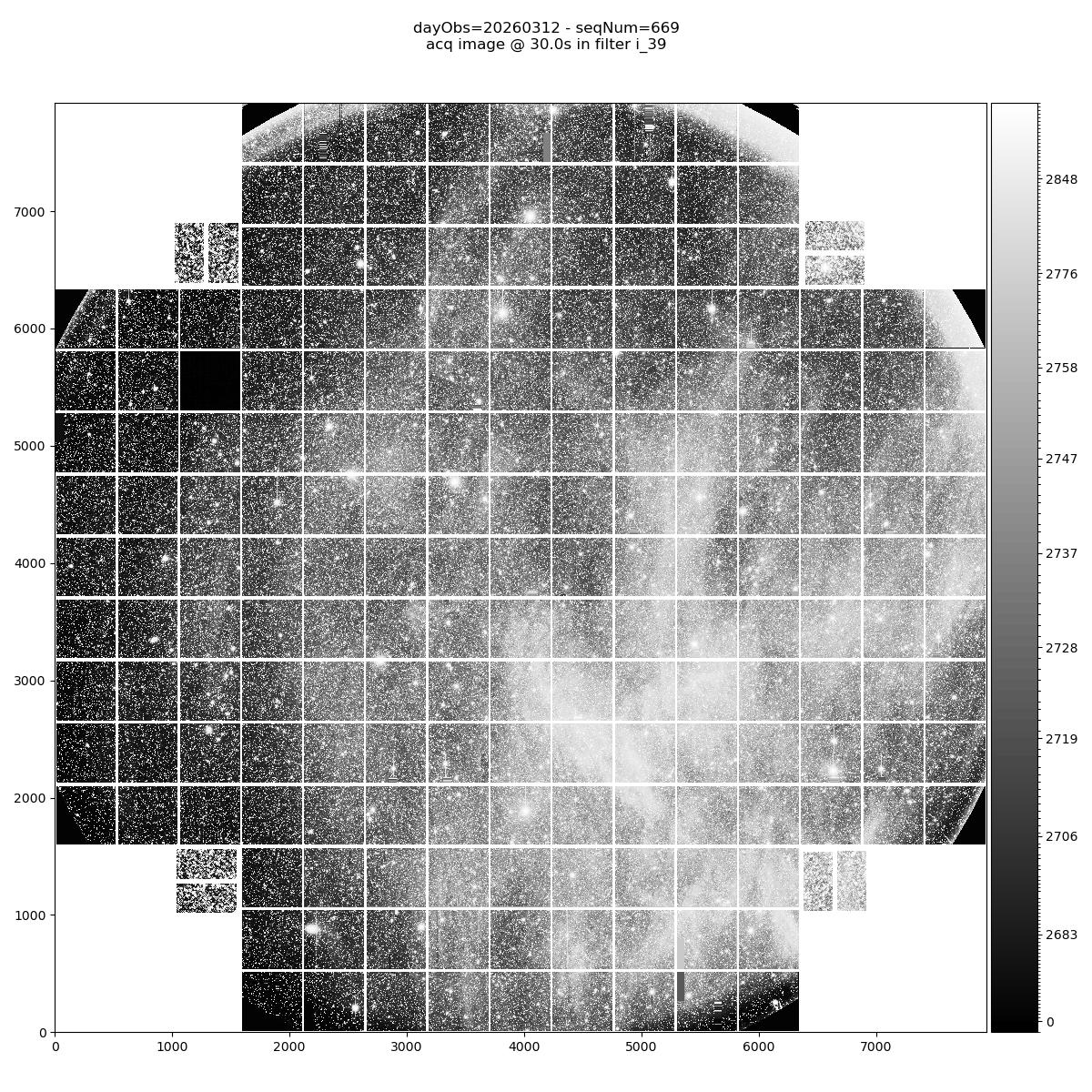}
    \caption{\label{fig:after} Mitigation of the prominent scratched tape stray light feature by the mid-level light baffle extension. (Left) Stray light originating from $\alpha$\,Cen located $\sim$21 degrees off-axis prior to the installation of the mid-level baffle extension (visit = 2026021600550; 30\,s exposure in $i$-band). (Right) Stray light is not present in an exposure taken after installation of the mid-level baffle extension (visit = 2026031200669; 30\,s exposure in $i$-band). The relative position of $\alpha$\,Cen is similar to the image on the left.}
    \vspace{1em}
    \includegraphics[width=1.0\textwidth, trim={1cm 1cm 1cm 0cm}, clip]{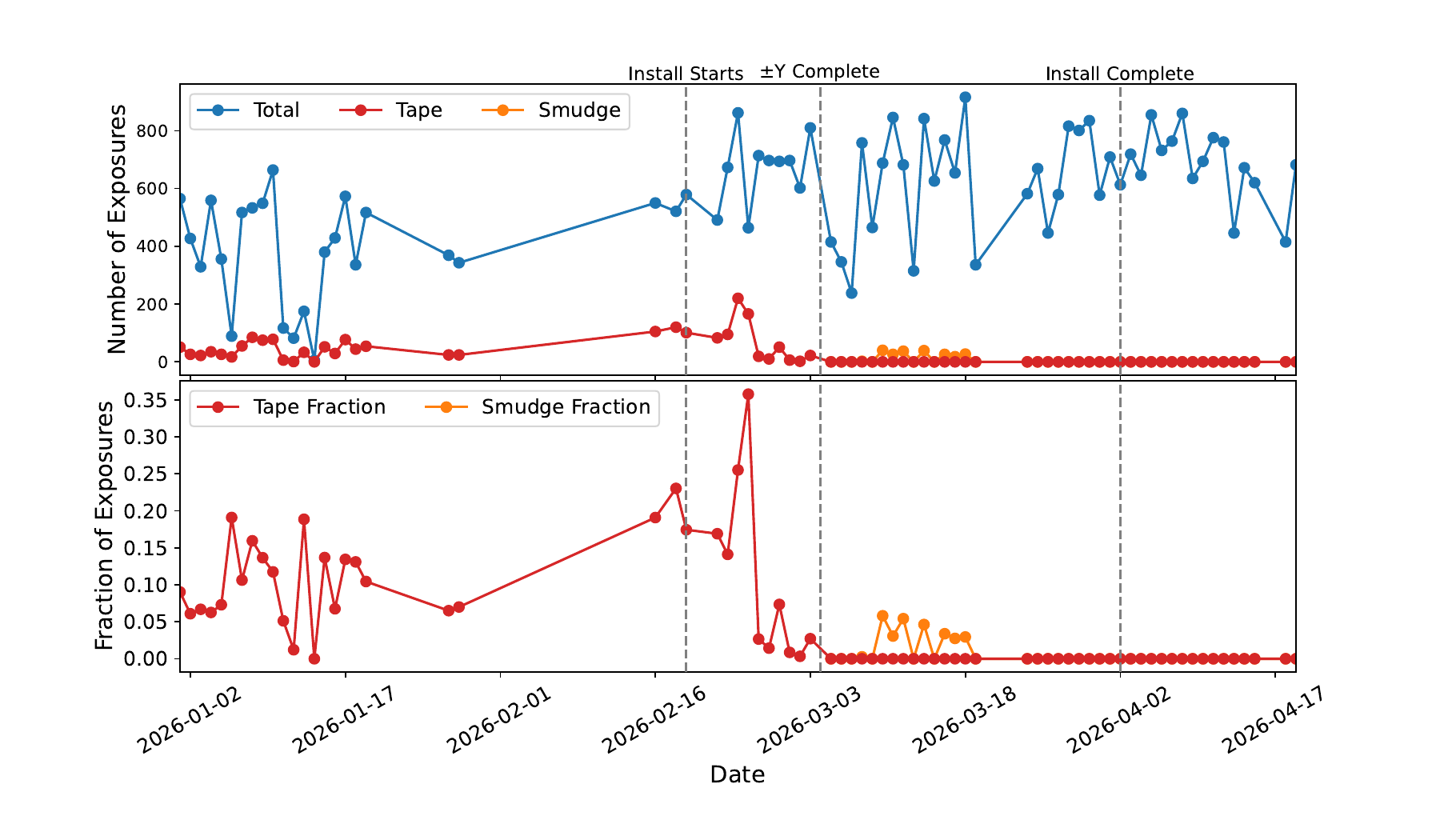}
    \caption{\label{fig:results} Installation of the mid-level light baffle extension effectively removed the scratched tape stray light feature. A small, residual feature (known as the ``smudge'') appeared briefly, and was attributed to a remaining gap between the extension panels. This feature has not been observed since installation was completed.}
\end{figure*}

The mechanical design of the mid-level baffle extension was performed at NOIRLab.
Consideration was given to a lightweight modular design that would allow installation to be staged during regular daytime engineering to avoid interrupting the observing schedule.
The baffle extension comprises 24 panels, each measuring 120\,cm $\times$ 27\,cm, allowing for 5\,cm of radial overlap with the existing mid-level baffle (Figure~\ref{fig:baffle} and \ref{fig:install}). 
The baffle extension panels were fabricated from 2-mm-thick low-carbon steel and painted with
low-reflectivity Aeroglaze Z306. 
These panels were attached to the mid-level baffle using self-tapping screws that were fixed in place with silicone sealant. Small overlap panels (22\,cm $\times$ 5\,cm) were attached to the extension panels to prevent gaps between the panels.

Installation of the mid-level baffle extension occurred over a period of $\sim$6 weeks between late February and early April, 2026 (Table~\ref{tab:timeline}).
The $\pm$Y quadrants have the largest scientific impact, since these quadrants allow a direct path to the sky, while the $\pm$X quadrants are already blocked by the dome.
In contrast, access to the $-$Y quadrant was the easiest, while access to the $\pm$X quadrants required a lift, and installation of the $+$Y quadrant required a section of the mid-level baffle to be removed from the TMA.
Installation generally progressed from the easiest sections to the most difficult, and installation of the main panels was completed in $\sim$\,2 weeks. Final installation of small overlap panels at the corners where the mid-level baffle connects to the TMA pylons was completed about a month later.

\section{Mitigation Results}
\label{sec:results}

The effectiveness of the mid-level baffle extension was validated with dedicated on-sky tests and with opportunistic on-sky data taking. Dedicated pre-observations on 16 February 2026 placed $\alpha$\,Cen in different radial and azimuthal positions in the scratched tape gap. Subsets of these observations were then repeated on 19 February 2026 (after the $-$Y extension was installed), 26 February 2026 (after the outer $+$Y panels were installed), 6 March 2026 (after the $+$Y panels were completed), and 12 March 2026 (after several corner overlap panels were installed). These tests demonstrated that the scratched tape light path had been blocked (Figure~\ref{fig:after}). A small residual feature (referred to as the ``smudge'') was observed until 19 March 2026. This smudge has not been observed since additional corner overlap panels were installed. Figure~\ref{fig:results} shows the relative occurrence rate of the scratched tape and smudge before, during, and after the installation of the mid-level baffle extension. Anecdotally, the muddy shoe and pillow stray light artifacts have also been observed less frequently since the installation of the mid-level baffle extension. However, these features occur much less frequently than the scratched tape feature and require additional observations to fully characterize.

\section{Conclusions}
\label{sec:conclusions}

The scratched tape stray light feature was caused by illumination from astronomical sources at large (19.8 $\lesssim \theta \lesssim$ 26\,deg) off-axis angles. This light passed through a gap between the mid-level and center-section light baffles on the TMA, reflected off the primary mirror, and reached the focal plane of LSSTCam. These features were prominent during LSSTCam commissioning and early operations. Visual inspection determined that bright instances of the scratched tape feature contribute to $>$\,5\% of on-sky science images during commissioning, science validation, and early operations. Instances of these features arising from bright stars were found to have a surface brightness comparable to $\sim$\,20\% of the dark night sky background in the bluer bands. While the impact of this feature on high-level scientific data products is mitigated through aggressive sky background modeling and subtraction, some residual effects are expected to persist in Rubin Data Preview 2. Based on a detailed analysis and modeling of the scratched tape feature, an outward extension of the TMA mid-level light baffle was proposed, designed, constructed, and installed in early 2026. The mid-level baffle extension successfully blocks the scratched tape stray light path, considerably reducing stray light contamination in the LSSTCam images prior to the start of LSST.

\acknowledgments 

This material is based upon work supported in part by the National Science Foundation through Cooperative Agreements AST-1258333 and AST-2241526 and Cooperative Support Agreements AST-1202910 and 2211468 managed by the Association of Universities for Research in Astronomy (AURA), and the Department of Energy under Contract No. DE-AC02-76SF00515 with the SLAC National Accelerator Laboratory managed by Stanford University. Additional Rubin Observatory funding comes from private donations, grants to universities, and in-kind support from LSST-DA Institutional Members.

The work of the authors was partially supported by the National Science Foundation Astronomy and Astrophysics Research Grants program through awards AST-2006340, AST-2307126, and AST-2407526. 

This document was prepared using the resources of the Fermi National Accelerator Laboratory (Fermilab), a U.S. Department of Energy, Office of Science, Office of High Energy Physics HEP User Facility. Fermilab is managed by Fermi Forward Discovery Group, LLC, acting under Contract No.\ 89243024CSC000002.

\bibliography{report} 
\bibliographystyle{spiebib} 

\end{document}